\documentclass{aastex}
\usepackage{spr-astr-addons}
\usepackage{url}
\usepackage{graphicx}
\usepackage{amsmath}
\usepackage{amsfonts}
\usepackage{amsthm}
\usepackage{float}
\usepackage[utf8x]{inputenc} 

\RequirePackage{color}

\begin{document}

\setlength{\overfullrule}{0pt}

\title{New Cutoff Frequency for Torsional Alf\'ven Waves 
Propagating along Wide Solar Magnetic flux Tubes}
%\slugcomment{Not to appear in Nonlearned J., 45.}
%% Running heads
\shorttitle{Cutoff for Alf\'ven Waves in Wide flux Tubes }
\shortauthors{Routh et al.}

\author{Swati Routh\altaffilmark{1}} 
\author{Z.E. Musielak\altaffilmark{2,3}}
\author{M.N. Sundar\altaffilmark{1}}
\author{Sai Sravanthi Joshi\altaffilmark{1}}
\author{Sree Charan\altaffilmark{1}}

\altaffiltext{1}{Department of Physics, School of Sciences, \\Jain University,Bengaluru, India.\\ 
email: routhswati@gmail.com}
\altaffiltext{2}{Department of Physics, University of Texas at Arlington,
       \\Arlington, TX 76019, USA.\\
       email: zmusielak@uta.edu}
\altaffiltext{3}{Leibniz-Institut f\"ur Sonnenphysik (KIS), Sch\"oneckstr. 6,
       79104 Freiburg, Germany.}

\begin{abstract}
An isolated, isothermal, and wide magnetic flux tube embedded either in the solar 
chromosphere or in the lower solar corona is considered, and the propagation of 
linear torsional Alfv\'en waves is investigated.  It is shown that the wideness of 
the tube leads to a new cutoff frequency, which is a local quantity that gives the 
conditions for the wave propagation at different atmospheric heights.  The cutoff 
is used to establish the ranges of frequencies for the propagating and reflected 
waves in the solar chromosphere and lower solar corona.  The obtained results 
are compared to those previously obtained for thin magnetic flux tubes and the 
differences are discussed.  Moreover, the results are also compared to some current 
observational data, and used to establish the presence of propagating waves in the 
data at different atmospheric heights; this has profound implications on the energy 
and momentum transfer by the waves in the solar atmosphere, and the role of linear 
torsional Alfv\'en waves in the atmospheric heating and wind acceleration.
\end{abstract}


%\section*{}
%\label{sec:intro}
\section{Introduction}

\par 
The presence of magnetic flux tubes in the solar atmosphere is well-established observationally.  
In the solar photosphere and lower chromosphere, these tubes can be approximated as thin, 
however, they must be considered as wide tubes in the upper chromosphere, transition region 
and solar corona (e.g.,\cite{1982Obs...102..118P, 1993SSRv...63....1S}).  The fundamental 
difference between the thin and wide magnetic flux tubes is that for the former the field remains 
the same for all magnetic field lines, however, it changes from one magnteic line to another in 
wide flux tubes (e.g.,\citealt{1997LNP...489...75R}).  Thin solar magnetic tubes can support 
longitudinal (sausage), transverse (kink) and torsional (Alfv\'en) waves and also fluting modes 
(e.g., \citealt{1982SoPh...75....3S, 1990JGR....9514873H, 1991GApFD..62...83R, 1997LNP...489...75R, 2004A&A...415..751S}).  Similar types of waves also exist in wide solar magnetic 
flux tubes but their propagation conditions are different (e.g., \citealt{1997LNP...489...75R}). 
The fact that different tube waves may significantly contribute to the local heating required in 
different layers of the solar atmosphere has been extensively investigated in the literature 
(e.g., \citealt{1996SSRv...75..453N,1997LNP...489...75R, 2003ASPC..286..363U, 2004IAUS..219..437M}).  

A significant amount of work was done on torsional Alf\'ven waves as they may carry enough 
wave energy and momentum to heat the upper solar chromosphere, transition region and corona 
(e.g., \citealt{1979ApJ...230..905P, 1980JGR....85.1311H,1982Obs...102..118P, 1985aspp.conf...77H, 1985A&A...151...16P, 1989A&A...210..425F, 1994JGR....9923489K, 1994ApJ...433..852F, 1997LNP...489...75R, 2007ApJ...659..650M, 2007SoPh..246..133R, 2010A&A...518A..37M, 2010ApJ...709.1297R, 2012JGRA..117.5229W, 2013MNRAS.428...40C, 2014ApJ...788....8M, 2015MNRAS.450.3169P}).  In the cited papers, thin solar magnetic flux tubes were 
mainly considered and the propagation conditions for torsional Alfv\'en waves were investigated.
Our approach presented in this paper is different as we deal primarily with wide solar magnetic 
flux tubes located either in the solar chromopshere or in the lower part of the solar corona; 
however, we validate our results by taking the limit of thin magnetic flux tubes and comparing 
the results in this limit with those presented in some of the above cited papers.

In general, the problem of propagation of torsional Alfv\'en waves along thin solar magnetic flux
tubes was formulated and studied in \cite{1978SoPh...56..305H,1981SoPh...70...25H,1982SoPh...75...79H}.  The fundamental physical concept in this problem is a cutoff frequency, which exists 
because of the presence of gradients of physical parameters in the solar atmosphere.  According 
to \cite{2007ApJ...659..650M}, the propagation of torsional Alfv\'en waves along thin and isothermal 
solar magnetic flux tubes is cutoff free.  However, the cutoff frequency also appears as a result of 
solar temperature gradients \cite{2007SoPh..246..133R}.   With the presence of inhomogeneities 
in the solar atmosphere, a new concept of cutoff frequencies is required because the cutoff originally 
introduced by Lamb (1911) can only be applied to a homogeneous atmosphere, which is unrealistic 
for the Sun as shown by \cite{2010ApJ...709.1297R}, \cite{2010A&A...518A..37M}, and 
\cite{2015MNRAS.450.3169P}.  

There are basic differences in settings between the papers by \cite{2007ApJ...659..650M} and 
\cite{2007SoPh..246..133R,2010ApJ...709.1297R}, who considered the propagation of linear 
torsional Alfv\'en waves along thin solar magnetic flux tubes, and the approach presented in 
this paper, which concentrates on linear torsional Alfv\'en waves and their propagation along 
wide solar magnetic flux tubes.  It must be also pointed out that in the work by \cite{2010A&A...518A..37M} and \cite{2015MNRAS.450.3169P}, linear transverse Alfv\'en waves were 
considered and their propagation along uniform magnetic field lines was investigated.  In all 
these papers, the cutoff frequencies were obtained, thus, our aim is to compare their results to
ours presented in this paper, which would allow us to determine how much the tube wideness 
affects the wave cutoff frequencies. 

We also compare our results to those previously obtained in numericla studies of dissipation and 
momentum deposition by torsional Alfv\'en waves propagating along thin solar magnetic flux 
tubes; the governing equations for those studies were derived by \cite{1981SoPh...70...25H,1982SoPh...75...79H} and then extended to more than one dimension with nonlinear terms included (e.g., \citealt{
1999ApJ...514..493K,2001ApJ...554.1151S,2013MNRAS.428...40C,2015SoPh..290.1909M, 2017SoPh..292...31W}).  Since most of this cited work concerns nonlinear waves, the comparison to our
linear results is limited to only several special cases.   We also plan to compare our results to some
observational data, so let us now briefly review these observations.

Observational evidence for the presence of torsional Alfv\'en waves in the solar atmosphere was given 
by\cite{2009Sci...323.1582J}, who analyzed high spatial resolution of H$\alpha$ observations by the
Swedish Solar Telescope (SST).  They identified Alfv\'en waves with periods from 12 min to 2 min and
with the maximum power near 6-7 min.  According to these authors, the waves carry enough energy to 
heat the solar corona, which is still questionable (\citealt{2010CSci...98..295D}).  Pure Alfv\'en motions 
were also observed by \cite{2008ApJ...687L.131B} who showed evidence for vortex motions of G band 
bright points around downflow zones in the photosphere; the lifetimes of these motions is aorund 5 min. 
In addition, \cite{2009A&A...507L...9W} demonstrated the presence of some disorganized relative motions 
of photospheric bright points and identified them as swirl-like motions in the solar chromosphere.   Finally, 
\cite{2011Natur.475..477M} reported indirect evidence for Alfv\'en waves found in observations by Solar 
Dynamic Observatory (SDO), see also \cite{2011Natur.475..463C}.  

Observations of atmospheric oscillations were made by SOHO and TRACE.   Periods of these oscillations 
range from 6 to 10 min, from 3 to 7 min and from 1 to 10 min respectively in the solar chromosphere, 
transition region and corona, respectively (e.g.,\cite{2003ApJ...595L..63D, 2005ApJ...624L..61D, 2004ApJ...609L..95M, 2004AAS...204.9806M, 2008AdSpR..42...86H}).  More recently, the GREGOR Infrared 
Spectrograph measured variations of the magneto-acoustic cutoff frequency in a sunspot umbra as 
reported by \cite{2018A&A...617A..39F} and found this cutoff to be of the order of $3.1$ mHz, which 
is in agreement with theoretical predictions.  The magneto-acoustic cutoff represents typical frequency 
of solar chromospheric oscillations and confirms that magnetoacoustic waves are responsible for the 
origin of these oscillations.  Since linear torsional Alfv\'en waves cannot excite these oscillations, the 
results obtained in this paper can only be compared to the available solar observations of Alfv\'en waves.

It was suggested that some transverse oscillations observed in the solar photosphere and chromosphere 
could be driven by Alfv\'en or kink waves (\cite{2008ApJ...676L..73V}).  Solar p-mode oscillations were 
also detected in sunspots (Zirin and Stein 1972), whose atmosphere with large and strong magnetic fields
maybe suitable for the propagation of Alfv\'en waves.  A mechanism of mode conversion from fast to slow 
modes in the lower chromosphere may also generate Alfv\'en waves (\cite{2011ApJ...738..119C}).  
Observations also show that the magnetic field over the umbra region of a sunspot reduces the amplitude 
of oscillations in the solar photosphere.  \cite{1992ASIC..375..261L} classified the waves in the sunspots 
into the following 3 categories: 5 minute oscillations in the sunspot photosphere, 3 minute oscillations and 
umbral flashes in the umbral chromosphere, and running penumbral waves in the penumbral chromosphere. 
It was found that the umbral waves and flashes travel upward along the vertical magnetic field in the umbral 
region of sunspots.  Optical observations of the solar chromosphere demonstrated that there is a connection 
between the umbral disturbances and penumbral waves (\cite{1982ApJ...253..386L}).   Finally, observations 
and polarimetric studies of the solar corona (\cite{2007Sci...318.1574D,2007Sci...317.1192T}) showed the 
presence of Alfv\'en waves.  

The main purpose of this paper is to determine analytically a new cutoff frequency and used it to determine the 
conditions for propagation of linear torsional Alfv\'en waves along isolated and isothermal magnetic flux tubes 
that are considered here to be wide.   As the obtained results show the derived cutoff is different than others 
previously obtained for thin magnetic flux tubes; the main reason is that for thin tubes there is no structure in 
the horizontal direction as all magnetic-field lines have the same physical properties across these tube.   Since 
at each given height of a wide flux tube, the magnetic field line is characterized by different physical parameters, 
different Alfv\'en velocity are obtained (e.g., \cite{1981SoPh...70...25H}). In addition, there is a gradient of 
Alfv\'en velocity along each field line.   Both effects lead to a new cutoff frequency that is determined here by 
using a method previously developed by \cite{2007ApJ...659..650M}.   The results obtained in this paper 
significantly generalize the previous work on cutoff frequencies in thin magnetic flux tubes (see citations 
above).   The theoretically established cutoff frequencies are compared to some available observational 
solar data as well as to some previously found cutoffs for Alfv\'en waves propagating in the solar 
atmosphere.   We use the comparison to discuss the role of the new cutoff in transporting energy 
and momentum by torsional Alfv\'en waves to different layers of the solar atmosphere.

The outline of our paper is as follows: in Section 2, we describe our flux tube model and present the basic equations. 
The cutoff frequency for torsional waves propagating along wide and isothermal magnetic flux tubes is derived in 
Section 3. The conditions for the wave propagation in the solar atmosphere are presented and discussed in Section 
4. Comparison of our theoretical results to the observational data is given in Section 5, and our conclusions are 
given in Section 6. 

\section{Governing and wave equations}
\par 
Following \cite{2007SoPh..246..133R}, we consider an isolated wide magnetic flux tube that is embedded 
either in the solar chromosphere or in the lower solar corona.   In our model, the solar atmosphere is approximated by an incompressible and isothermal medium that has fixed density stratification.   The 
tube is untwisted, has a circular cross-section and is in temperature equilibrium with the external medium.   According to \cite{1978SoPh...56..305H, 1981SoPh...70...25H} (also \citealt{1999ApJ...514..493K, 2001ApJ...554.1151S}), the propagation of linear torsional Alfv\'en waves along such a flux tube can be described by 
using an orthogonal curvilinear coordinate system $(\xi,\theta, s)$, where s is a parameter along a given magnetic field line, $\theta$ is the azimuthal angle about the axis of symmetry, and $\xi$ is a coordinate 
in the direction $\hat{\xi} = \hat{\theta} \times \hat{s}$.

\par 
The background magnetic field becomes $B_{0s} = B_{0s} (s)\hat{s}$
with $B_{0 \xi}=0$ and $B_{0 \theta}=0$. Since only linear torsional waves are considered, the pressure and density variations associated with the waves are neglected, and the waves are described by $v=v_\theta (s,t)\theta$ and $b=b_\theta (s,t)\theta$.

\par 
From the work of \cite{1982SoPh...75...79H}, the curvilinear scale factors, $h_{\theta}=R$ , are introduced, where $R=R(s)$ represents the distance from the magnetic-field line to the tube axis, $h_s=1$ , and $h_\xi=R$, with the latter being determined by the conservation of magnetic flux. Using these scale factors, we obtain the following momentum and induction equations

\begin{equation}
\label{basic1}
\frac{\partial}{\partial t} \left( \frac{v_{\theta}}{R} \right) - 
\frac{B_{0s}}{4 \pi \rho_{0} R^{2}}  \frac{\partial}{\partial t} 
\left( Rb_{\theta} \right) = 0\ ,
\end{equation}

\begin{equation}
\label{basic2}
\left( Rb_{\theta} \right) -R^{2}B \frac{\partial}{\partial s}
\left( \frac{v_\theta}{R} \right) = 0\ .
\end{equation}

These are the basic equations that describe the propagation of torsional waves along the magnetic flux tube embedded in the solar atmosphere. 

We combine (\ref{basic1}) and (\ref{basic2}) and derive the wave equations for torsional tube waves. For this we consider the following two sets of the wave variables: $v_\theta$ and $b_\theta$, and $x \equiv \frac {v_{\theta}}{R} $ and $y \equiv Rb_{\theta}$, which are used here to study the wave behaviour in the solar atmosphere \cite{1982SoPh...75...79H}.

To derive the wave equations for the variables $v_\theta$ and $b_\theta$, we use the conservation of magnetic flux $[ \pi R^{2}(s) B_{0s} = constant ]$ to express $R(s)$ in terms of $B_{0s}$, and applying them in MHD equations (\ref{basic1} , \ref{basic2}) gives the wave equations as: \\

%\begin{align}
\label{wave1}
$\frac{\partial^{2} v_{\theta}}{\partial t^{2}} - C_{A}^{2}\frac{\partial^{2} v_{\theta}}{\partial s^{2}}-\frac{C_{A}^{2}}{B_{0s}} \left( \frac{dB_{0s}}{ds} \right) \frac{\partial  v_{\theta}}{\partial s} + \nonumber \\*
\\C_{A}^{2} \left[ \frac{1}{4B_{0s}^{2}} \left( \frac{dB_{0s}}{ds} \right) ^{2} - \frac{1}{2B_{0s}} \left( \frac{d^{2B_{os}}}{ds^{2}}\right) \right] v_{\theta}=0$\\
%\end{align}

%\begin{align}
\label{wave2}

$\frac{\partial^{2} b_{\theta}}{\partial t^{2}} - C_{A}^{2} \frac{\partial^{2} b_{\theta}}{\partial s^{2}} + C_{A}^{2} \left[ \frac{1}{B_{os}} \left( \frac{dB_{0s}}{ds} \right) - \frac{2}{C_{A}} \left( \frac{dC_{A}}{ds} \right) \right] \frac{\partial b_{\theta}}{\partial s} +  \nonumber \\
 \\C_{A}^{2} \left[ \frac{1}{C_{A}}  \frac{dB_{0s}}{ds} - \frac{3}{4B_{os}^{2}}\left( \frac{dB_{0s}}{ds} \right) ^{2} \right] + \frac{1}{2B_{os}}\left( \frac{d^{2}B_{0s}}{ds^{2}} \right) b_{\theta} = 0$\\
%\end{align}
where $C_{A}(s)$=$\frac{B_{os}(s)}{\sqrt[2]{4 \pi \rho_{0}(s)} }$ is the Alfv\'en velocity along a given magnetic-field line. Note that the derived wave equations have different forms, which means that the behavior of the wave variables $v_{\theta}$ 
and $b_{\theta}$ is not the same.  Similarly, using the Hollweg variables $x$ and $y$, Eqs (\ref{basic1}) and (\ref{basic2}) 
give also two different wave equations
\begin{equation}
\label{wave3}
\frac{\partial^{2} x}{\partial t^{2}} -C_{A}^{2}\frac{ \partial^{2} x }{ \partial s^{2} }  = 0\ ,
\end{equation}
and
\begin{equation}
\label{wave4}
\frac{\partial y^2}{\partial t^{2}} - \frac{\partial }{\partial s} \left[ C_{A}^{2}(s)\frac{\partial y}{\partial s}\right] = 0\ ,\\
\end{equation}
where the condition $B_{0s}R^{2}(s) = constant$ is satisfied.

The difference in the derived wave equations reflect the fact that $x$ and $y$ behave differently. Comparison of these wave equations to those derived for the wave variables ($v_{\theta}$ and $b_{\theta}$) clearly shows that each wave variable has different behavior. The comparison also shows that there is an advantage in using Hollweg's variables because the wave equations for these variables are much simpler than those obtained for the variables $v$ and $b$. An interesting result is that the first set of wave equations derived above, is of the same form as that obtained by Musielak et al. (2006) in their studies of acoustic waves propagating in a non-isothermal medium. 

\section{Cutoff frequency: theory}

Musielak et al. (2006) developed a method to determine a cutoff frequency from given wave equations. The first step of this method is to transform the wave equations into the corresponding Klien-Gordon equations and then the cutoff frequency is derived by using the oscillation theorem (e.g., Kahn, 1990). Let us introduce the new variable $d \tau = \frac{ds}{C_{A}}$ and following \cite{2007SoPh..246..133R}, we write the critical frequencies as
\begin{equation}
\label{kg2}
\Omega^{2}_{x,v_{\theta}}(\tau) = \frac{3}{4}\left( \frac{C'_{A}}{C_{A}} \right) ^{2} - \frac{1}{2}\left( \frac{C''_{A}}{C_{A}} \right)\ , 
\end{equation}
and
\begin{equation}
\label{kg3}
\Omega^{2}_{y,b_{\theta}}(\tau) = \frac{1}{2}\left( \frac{C''_{A}}{C_{A}} \right)  - \frac{1}{4}\left( \frac{C'_{A}}{C_{A}} \right) ^{2} \ ,\\
\end{equation}
where $C'_{A} = \frac{dC_{A}}{d\tau}$ and $C''_{A} = \frac{d^{2}C_{A}}{d\tau^{2}}$.  Using these frequencies,
we define their corresponding turning-point frequencies $\Omega_{tp}$ as
 \begin{equation}
 \label{tp1}
 \Omega^{2}_{tp} =  \Omega^{2}_{crit} + 1/4\tau^2\ ,\\
 \end{equation}
where $\Omega_{crit}$ is either $\Omega_{x,v_{\theta}}$ or $\Omega_{y,b_{\theta}}$.

These turning-point frequencies separate the solutions into propagating and non-propagating (evanescent) waves. Since there is the turning-point frequency for each wave variable, only one of them can be the cutoff frequency. We follow \cite{2007ApJ...659..650M} and identify the largest turning-point frequency as the cutoff frequency. The choice is physically justified by the fact that in order to have propagating torsional tube waves, the wave frequency $\omega$ must always be higher than any turning-point frequency. To determine which turning-point frequency is larger, we need to specify a model of magnetic flux tubes embedded in the solar atmosphere.

A single and isothermal magnetic-flux tube is considered to be wide when its horizontal magnetic field is nonuniform, which means that each magnetic-field line has different physical properties in the horizontal direction. Let us assume that this tube is approximated by a simple model in which the Alfv\'en velocity varies exponentially along a given field line; the model was originally considered by \cite{1981SoPh...70...25H}, and it shall be used here to determine the cutoff frequency for torsional tube waves propagating in this model.

For the {\it exponential model}, the Alfv\'en velocity is given by 

\begin{equation}
\label{tor1}
C_{A} = C_{A0} e^{\left( \frac{s}{mH} \right)}\ .
\end{equation}
where $C_{A0} = C_{A}$ at $(s=0)$ , $m$ is a positive scaling factor, and $h$ is the characteristic 
scale height; we take $h = H$, with $H$ being the pressure (density) scale height. The reason for 
choosing different values of $m$ is that, in general, $C_{A}(s)$ must be different for each magnetic 
field line.  To calculate $\tau(s)$ we evaluate the following integral
\begin{equation}
\label{tor2}
\int_{\tau_0 }^{\tau} d\tau = \int_{0 }^{s} \frac{1}{C_{A}} ds\ ,
\end{equation}
which, after substitution and integration, becomes
\begin{equation}
\label{tor3}
\tau = - mH \left[ \frac{1}{C_{A}} - \frac{1}{C_{A0}}\right]\ +\tau_{0}\ .
\end{equation}
Thus, the characteristic wave velocity $C_{A}$ is expressed as,
\begin{equation}
\label{tor4}
C_{A} = \frac{mHC_{A0}}{mH - (\tau-\tau_{0})C_{A0}}
\end{equation}

Differentiating $C_{A}$ with respect to $\tau$ twice we respectively obtain
\begin{equation}
\label{tor5}
\frac{ d C_{A} }{d\tau}=\frac{ mH\ C_{A0}^2 }{ [\ mH - (\tau - \tau_{0}) C_{A0}  ]\ ^{2} }\ ,
\end{equation}
and
\begin{equation}
\label{tor6}
\frac{ d^{2} C_{A} }{d\tau^2}=\frac{ 2mH\ C_{A0}^3 }{ (\ mH - (\tau - \tau_{0}) C_{A0}  )\ ^{3} }\ ,
\end{equation}

Substituting equations (~\ref{tor4}),(~\ref{tor5}) and (~\ref{tor6}) in (~\ref{kg2}) and (~\ref{kg3}), the obtained equations are,

\begin{equation}
\label{tor7}
\Omega^{2}_{x}(\tau) =- \frac{1}{4}  \frac{C_{A0}^{2} }{ \left( mH - (\tau - \tau_{0}) C_{A0} \right)^{2}}\ ,
\end{equation}
and
\begin{equation}
\label{tor8}
\Omega^{2}_{y}(\tau) = \frac{3}{4}  \frac{C_{A0}^{2} }{ \left( mH - (\tau - \tau_{0}) C_{A0} \right)^{2}}\ .
\end{equation}

The cutoff frequency is given by
\begin{equation}
\label{tor9}
\Omega^{2}_{cut}(\tau) = Max \left[ \Omega^{2}_{x}(\tau) , \Omega^{2}_{y}(\tau)  \right] + \frac{1}{4\tau ^{2}}\ ,
\end{equation}

which becomes

\begin{equation}
\label{tor10}
\Omega^{2}_{cut}(\tau) = \frac{3}{4}  \frac{ C_{A0}^{2} }{ \left( mH - 
\left( \tau - \tau_{0}\right) C_{A0}\right) ^{2}} +\frac{1}{4\tau ^{2}}\ .
\end{equation}

The cutoff frequency $\Omega^{2}_{cutoff}$ in terms of $s$ is given by
\[
\Omega^{2}_{cut}(s) = \frac{3}{4}\frac{C_{A0}^{2} e^{(\frac{2s}{mH}})} 
{m^{2} H^{2}} 
\]
\begin{equation}
\hskip0.25in + \frac{C_{A0}^{2}}{4 \left[ \tau_{0}C_{A0} - mH 
\left( e^{\left(\frac{-s}{mH}\right)} - 1 \right)\right] ^{2}}\ . 
\label{tor11}
\end{equation}

Clearly, the cutoff frequency is a local quantity that varies with s in the same way as $c_{A}$ does. Because $\Omega_{cut}$ depends on height, its physical meaning is different than the global cutoff frequencies for longitudinal and transverse tube waves obtained by \cite{1976ApJ...209..266D} and \cite{1982SoPh...75....3S}, respectively. From a physical point of view, $\Omega_{cut}(s)$ represents locally the cutoff in the solar atmosphere, and torsional tube waves must have their frequency ($\omega$) higher than $\Omega_{cut}$  at a given height to reach this height and be propagating waves at this height. In other words, the cutoff allows us to determine the height ($s$) in the model at which torsional Alfv\'en waves of a given frequency become non-propagating (evanescent) waves.

\section{Cutoff frequency: results}

The formulas for the cutoff frequencies derived in the previous section are now used to calculate these cutoffs in the models of magnetic flux tubes considered in this paper. The calculations are performed for a magnetic flux tube with $B_{0}$ = 1500 G at the atmospheric height $s = 0$, which corresponds to the location in the solar atmosphere where the flux tubes have widened enough so that $c_A$ is no longer constant but increases exponentially with height. This height depends on the local magnetic filling factor and typically it should lie at the solar temperature minimum level or in the lower chromosphere. 

\begin{figure}
\begin{center} 
\includegraphics[width=0.4\textwidth]{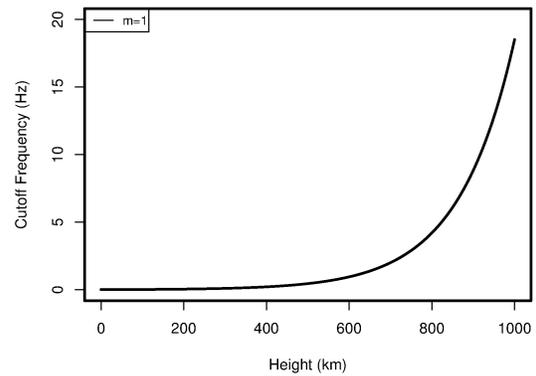}
\caption{The cutoff frequency $\Omega_{cut}$ plotted versus height for the exponential model with 
m=1 in the solar chromosphere with $C_{A0}$=11 km/s, $\tau_0$= 50 s, H=135 km.}
\end{center}
\end{figure}

\begin{figure}
\begin{center}
\includegraphics[width=0.4\textwidth]{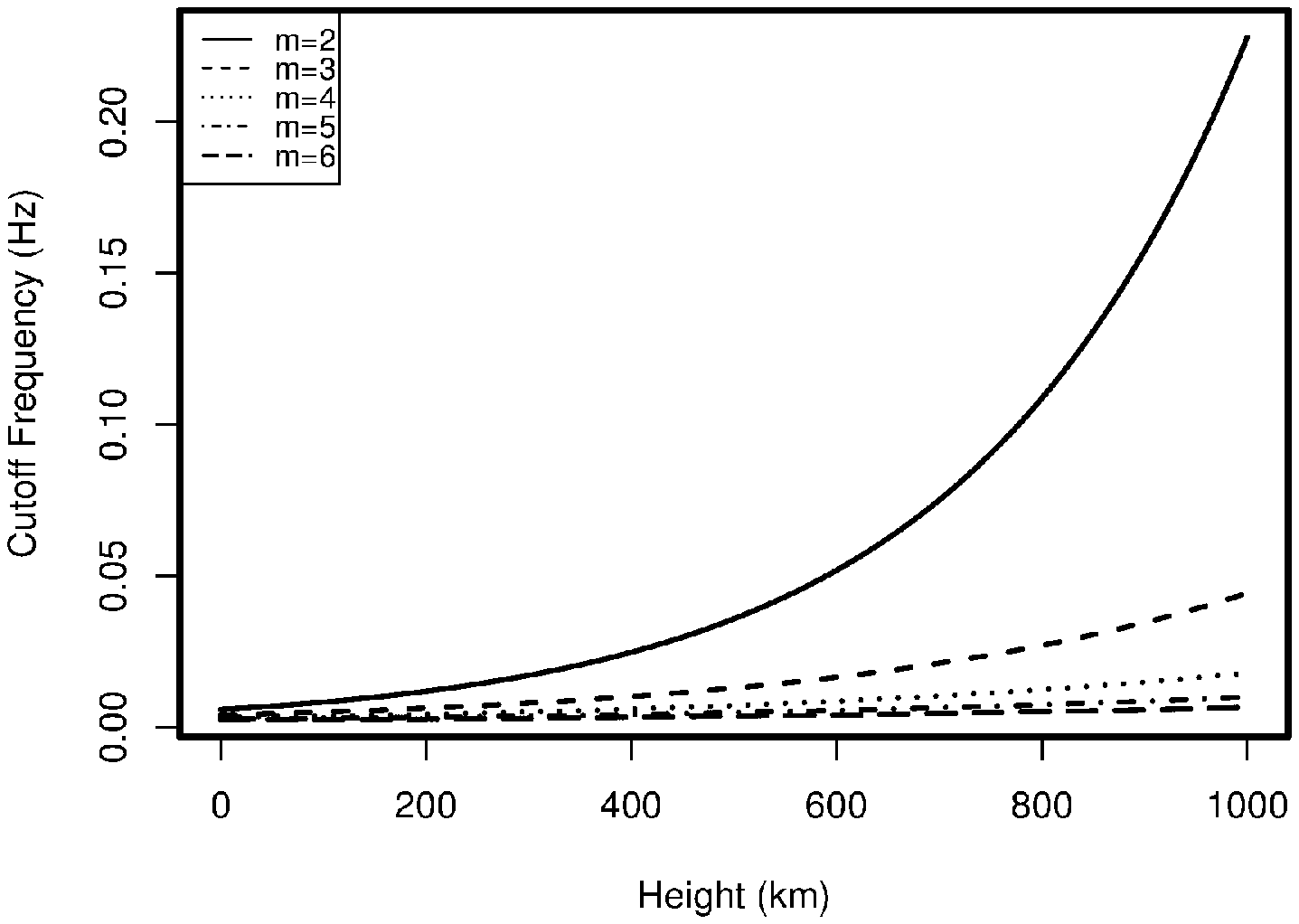}
\caption{The cutoff frequency $\Omega_{cut}$ plotted versus height for the exponential models with 
m=2 to m=6 in the solar chromosphere with $C_{A0}$=11km/s, $\tau_0$= 50 s, H=135 km.}
\end{center}
\end{figure}

As the characteristic temperature for our chromosphere model, we choose the effective temperature of the Sun ($T_{eff} =5770$ K).  The considered chromosphere model isothermal with $T_{eff}$, which gives $C_{A0} 
= 11.0 km/s$ and $H$ = 135 km; moreover, $\tau_0 = 50$ s, and the height $s = 0$ is located at the temperature minimum, more precisely 500 m above it.  The calculated cutoff frequencies are local quantities 
and their variations with height in the solar chromosphere are shown in Figs 1 and 2; in both figures the 
values of $\Omega_{cut} / 2 \pi$ are plotted.  As expected, the cutoff is much steeper for low values of 
$m$. It is also seen that the effect of the cutoff on the wave propagation becomes important at atmospheric 
heights higher than 100 km above the base of the model. 

The solar corona model considered in this paper is also isothermal with the temperature
$200 \times T_{eff}$, which gives  $C_{A0} = 1.05 Mm/s$ and $H$ = 60 Mm; these 
values are the same as those used by \cite{2010A&A...518A..37M}.  Moreover, $\tau_0 
= 200$ s and for the coronal models the height $s = 0$ is located above the solar 
chromosphere and corresponds 10 Mm.   The variations of the cutoff freqeuncies with 
height in the solar corona are shown in Figs 3 and 4 for different values of $m$; again, 
in both figures the values of $\Omega_{cut} / 2 \pi$ are plotted.   A steep 
increase of the cutoff frequency with height is seen for the models with m = 1 and m = 2.

For the torsional Alfv\'en waves to be propagating along a wide magnetic flux tube embedded in the solar atmosphere, it is required that the wave frequency $\omega$ exceeds the cutoff frequency $\Omega_{cut}$.  Specific values of the cutoff frequencies for different flux tube models are given in Table 1 and 2 for the solar chromosphere and corona, respectively; in both tables the values of $\Omega_{cut} / 2 \pi$ are plotted.
It must be noted that these values are given only for the bottom and top of each flux tube model label by different $m$.   The values of the cutoff frequencies range from 0.011 Hz  ($m=1)$ to 0.002 Hz ($m=5)$ 
at the bottom, and 0.23 Hz ($m=2)$ to 0.01 Hz ($m=5$) at the top of the solar chromosphere (see Table 1). However, according to Table 2, the cutoff period ranges from 0.0024 Hz ($m=1)$ to 0.0007 Hz ($m=4)$ 
at the bottom, and 0.0043 Hz ($m=1)$ to 0.0008 Hz ($m=4$) at the top of the solar corona.

\begin{figure}
\begin{center} 
\includegraphics[width=0.4\textwidth]{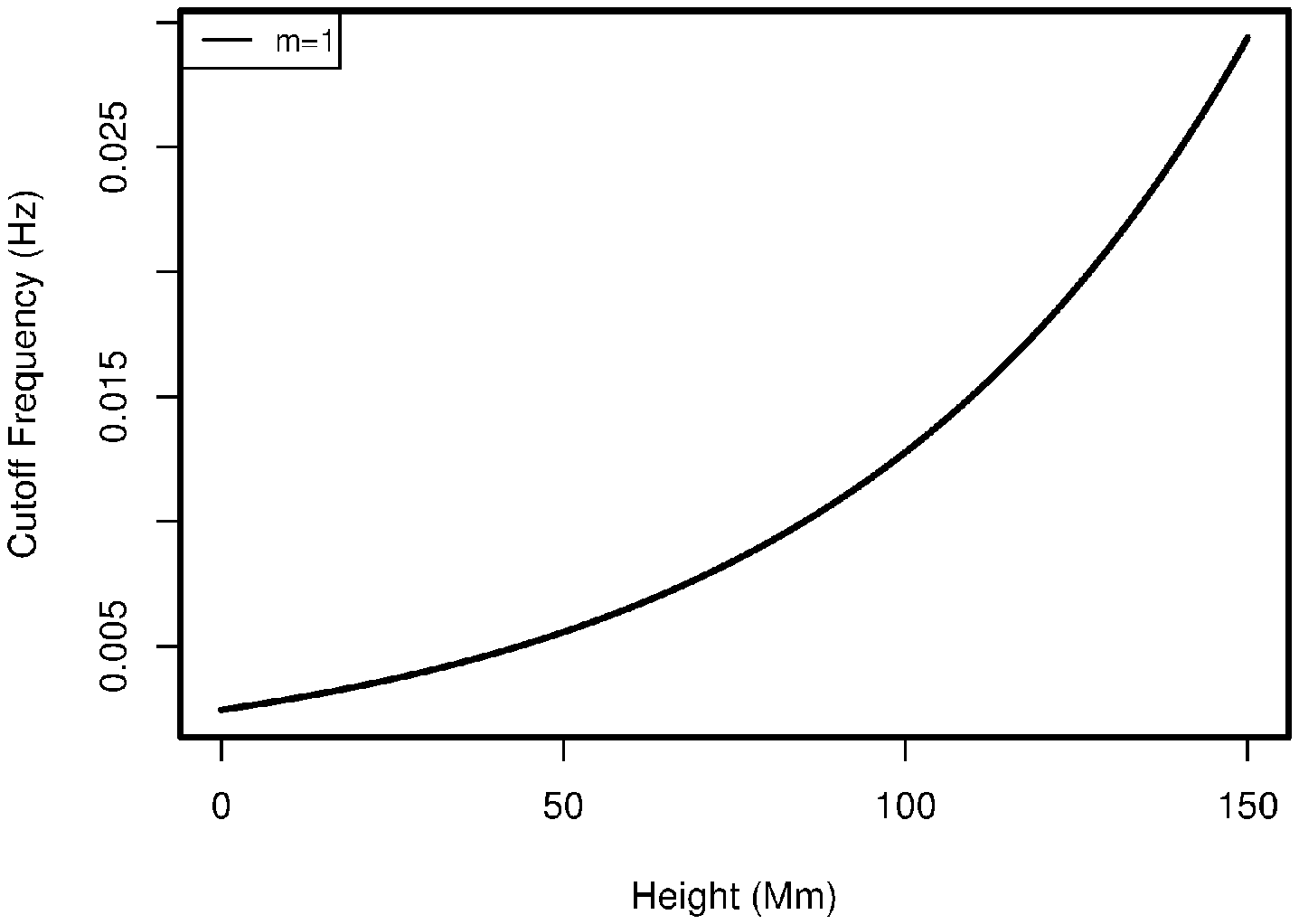}
\caption{The cutoff frequency $\Omega_{cut}$ plotted versus height for the exponential model 
with m=1 in the solar corona with $C_{A0}$=1.05Mm/s, $\tau_0$= 200 s, H=60 Mm.}
\end{center}
\end{figure}

\begin{figure}
\begin{center} 
\includegraphics[width=0.4\textwidth]{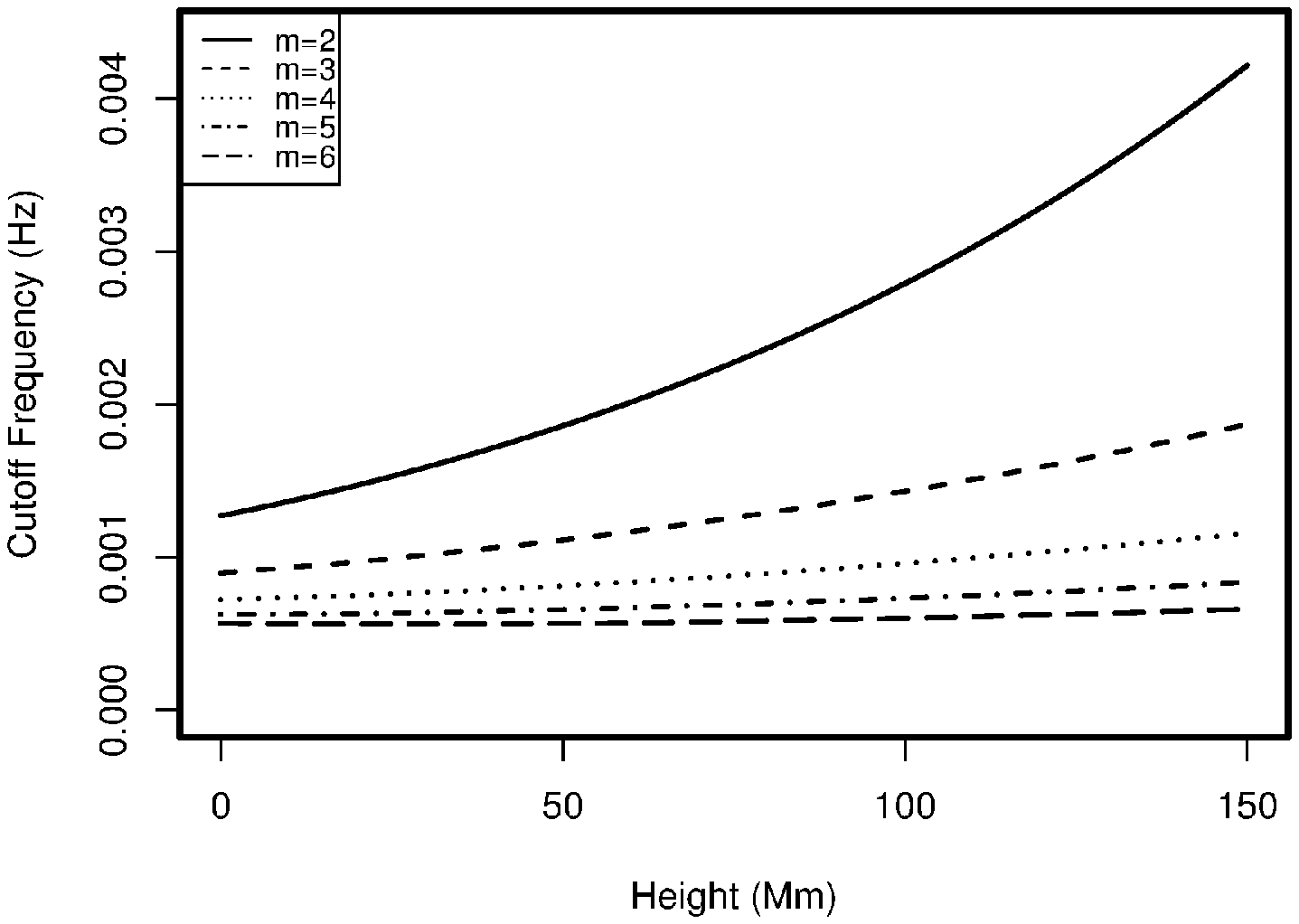}
\caption{The cutoff frequency $\Omega_{cut}$ plotted versus height for the exponential model 
with m=2 to m=6 in the solar corona with$C_{A0}$=1.05Mm/s, $\tau_0$= 200 sec, H=60 Mm.}
\end{center}
\end{figure}

The above cutoff frequencies for the solar chromosphere are consistent with the values of the cutoff frequencies for Alfv\'en waves propagating in the non-isothermal solar chromosphere with a uniform magnetic field studied by \cite{2010A&A...518A..37M}, whose cutoff frequencies range from 0.01 Hz at the bottom of their model to 
0.25 Hz at the top of their model.  Moreover, their value of the cutoff frequency in the solar corona is around 
0.01 Hz, which show significant differences with the results presented in Figs 3 and 4, and in Table 2.  These 
differences are likely to be the effects of magnetic fields, which in this paper is non-uniform and diverges rapidly with height but in \cite{2010A&A...518A..37M}'s paper the field is uniform. The values of cutoff periods obtained in this paper are also consistent with studies performed by \cite{2015MNRAS.450.3169P}, whose cutoff frequency ranges from 0.022 Hz to 0.005 Hz for propagating linear Alfv\'en waves in an isothermal solar atmosphere with a uniform magnetic field. 

\begin{table} [h]
\centering
\begin{tabular}{c	c	c}
\hline
\textbf{} & \textbf{Solar chromosphere}\\
\hline
\textbf{} & \textbf{Cutoff frequencies [Hz]}\\
\textbf{Model} & \textbf{Bottom} & \textbf{Top}\\
\hline
m = 1 & 0.011 & 18.51 \\
m = 2 & 0.006 & 0.228 \\
m = 3 & 0.004 & 0.044 \\
m = 4 & 0.003 & 0.018 \\
m = 5 & 0.002 & 0.010 \\
\hline
\end{tabular}
\caption{The cutoff frequencies of our flux tube models
with different values of $m$ calculated at the bottom 
and top layers of the solar chromosphere}
\end{table}

In our previous work on cutoff frequencies for transverse Alfv\'en waves in a thin non-isothermal magnetic flux tube (e.g., \citealt{2010ApJ...709.1297R, 2010AN....331..593H}), the ranges of cutoff periods for different regions of the solar chromosphere are: 0.003-0.008 Hz for the lower chromosphere, 0.003-0.008 Hz for the middle {chromosphere}, and 0.003-0.005 Hz for the upper chromosphere. Again, these results are consistent with those obtained in this paper, and some discrepancies between the results reflect the differences in the models of considered flux tubes, thin (previous work) and wide (this paper).

\begin{table}
\centering
\begin{tabular}{c	c	c}
\hline
\textbf{} & \textbf{Solar corona}\\
\hline
\textbf{} & \textbf{Cutoff frequencies [Hz]}\\
\textbf{Model} & \textbf{Bottom} & \textbf{Top}\\
\hline
m = 1 & 0.0024 & 0.0043\\
m = 2 & 0.0013 & 0.0017\\
m = 3 & 0.0008 & 0.0009\\
m = 4 & 0.0007 & 0.0008\\

\hline
\end{tabular}
\caption{The cutoff frequencies of the flux tube models
with different values of $m$ calculated at the bottom 
and top layers of the solar corona}
\end{table}

\section{Comparison to observations}

There is observational evidence for Alfv\'enic-like fluctuations in the solar corona and chromosphere (\cite{2007Sci...318.1574D, 2007Sci...317.1192T, 2008ApJ...687L.131B, 2009ApJ...702.1443F, 2009Sci...323.1582J,2009A&A...507L...9W}), and the observations show wave periods ranging from 3 to 9 min, which is consistent with the theoretical predictions of this paper. Moreover, \cite{2009ApJ...702.1443F} reported a detection of torsional tube waves with periods ranging from 0.001 Hz to 0.008 Hz, up to the top of the solar chromosphere in the expanding wave-guide rooted in the strongly magnetized MBP (magnetic bright point). Their observational results match our theoretical results in the solar chromosphere (especially for m=5, which gives period range as 0.002 Hz to 0.01 Hz) and corona (for m=1 and 2). In general, our results predict low frequency waves to be present in the solar corona while higher frequency waves are propagating in solar chromosphere, which may have implications on the energy and momentum transfer by linear torsional Alfv'en waves in the solar chromosphere and corona.  Another important aspect of our results is that they can be used to determine whether the observed waves are propagating or evanescent (e.g., Hammer, Musielak and Routh 2010).

\section{Conclusions}

We determined the conditions for propagation of linear (fully incompressible) torsional Alfv\'en waves along an isolated and isothermal magnetic flux tube embedded in the solar chromosphere and corona.  The main difference between the results of this paper and those considered previously (e.g., \cite{1981SoPh...70...25H, 1982SoPh...75...79H, 2007ApJ...659..650M, 2007SoPh..246..133R, 2010ApJ...709.1297R, 2010AN....331..593H}) is the fact that in the previous work thin magnetic flux tubes were studies but this paper investigates the effects caused by wide magnetic flux tubes on the torsional Alfv\'en wave propagation. It must be noted that in wide magnetic flux tubes the horizontal magnetic field is non uniform, and that the Alfv\'en velocity varies with height along the tubes.  

Our main theoretical result is derivation of the cutoff frequencies for linear torsional Alfv\'en waves propagating along wide magnetic flux tubes embedded in the solar chromosphere and corona. The cutoff frequency is a local quantity and its variation with height is used to identify regions in the solar atmosphere where strong wave reflection occurs. Using the condition $P_{w} = P_{cut}$, where $P_{w}$ is the wave period, the atmospheric height at which Alfv\'en waves of a given period are reflected can be determined. Wave reflection and the resulting constructive interference between the propagating and reflected Alfv\'en waves can form standing wave patterns as also seen in numerical studies performed by \cite{2010A&A...518A..37M}.

The theoretically predicted cutoff frequency for torsional Alfv\'en waves propagating along wide magnetic flux tubes is consistent with the previous studies performed by \cite{2010A&A...518A..37M}, \cite{2010ApJ...709.1297R}, \cite{2010AN....331..593H}, and \cite{2015MNRAS.450.3169P} in the solar chromosphere. 
However, the novelty of this work is that the computed cutoff frequencies in the solar corona are different 
than those found previously because in the wide magnetic flux tubes considered here, the magnetic field
alters greatly with the atmospheric height.  The derived chromospheric and coronal cutoffs are in agreement 
with the currently available observational data; actually our results can be used to determine whether the observed Alfv\'en waves are propagating or not in the solar chromosphere and corona.\\
\\
\\{\bf Acknowledgment} We are indebted to an anonymous referee for valuabe comments and suggestions 
that allow us to significantly improved our original manuscript.

\bibliographystyle{spr-mp-nameyear-cnd}
%\bibliography{biblio3}  

\end{document}